\documentstyle[prl,twocolumn,aps,epsf,epsfig]{revtex}
\begin{document}
\draft
\twocolumn[\hsize\textwidth\columnwidth\hsize\csname@twocolumnfalse%
\endcsname

%\documentstyle[prl,twocolumn,aps,epsf,epsfig]{revtex}
%\begin{document}
%\draft

\title{Effect of the Equivalence Between Topological and Electric Charge
on the Magnetization of the Hall Ferromagnet.}

\author{Luis Brey}
\address{Instituto de Ciencia de Materiales de Madrid, 
CSIC,
28049 Cantoblanco, Madrid, Spain.}

\date{\today}

\maketitle

\begin{abstract}
\baselineskip=2.5ex

The dependence on temperature of the spin magnetization of a 
two-dimensional electron gas at filling factor unity is studied.
Using classical Monte Carlo simulations we analyze the effect that the
equivalence between topological and electrical charge has on the
the behavior of the magnetization. We find that at 
intermediate temperatures the spin polarization increases in a thirty per cent
due to the Hartree
interaction between charge fluctuations.

\vspace{0.5truecm}
\end{abstract}
PACS number  73.40.Hm, 73.20Dx, 73.20Mf
]
%\tableofcontents
The physics of a two-dimensional electron gas (2DEG) in a magnetic field ,$B$, is determined
by the number of Landau levels occupied by the electrons. Since the degeneracy
of the Landau levels  increases
with $B$, all the electrons can be accommodated in the 
lowest Landau level for strong enough fields.
At filling factor unity, $\nu=1$, and zero temperature, $T=0$, the ground 
state of the 2DEG is an itinerant ferromagnet.\cite{ahm1,libro1}
The Zeeman coupling between the electron spin and the magnetic field
determines the orientation of the ferromagnet polarization. 
For zero Zeeman  coupling the interaction between the carriers produces
a spontaneous spin magnetic moment.

Recently the magnetization of the quantum Hall system was 
measured\cite{barret} for filling
factors $0.66< \nu < 1.76$ and temperatures $1.55K < T < 20K$.
At $\nu=1$ and very low temperatures the system is fully polarized, while for
other filling factors the magnetization is reduced. 
The demagnetization for $\nu \neq 1$ is related to 
the existence of Skyrmions in the system.\cite{theory1}
Further experiments have verified the existence of Skyrmions using 
transport, capacity  and optical experiments.\cite{otros,manfra}

For $\nu=1$, the temperature dependence of the magnetization $M(T)$,
has been measured using NMR techniques\cite{barret} and 
magneto-optical absorption experiments.\cite{manfra}
Different theoretical approaches have been used for the 
study of $M(T)$:
i) Read and Sachdev\cite{read} have studied the $N \rightarrow \infty$
limit of a quantum continuum field theory model for  the spin vector
field, ${\bf m} ( {\bf r} )$.
This model describes the long-wavelength collective behavior of the
electronic spins. This work has been extended by Timm {\it et al.}\cite{timm}
The field theory is expected to be accurate at low temperatures and   
weak Zeeman coupling. Using $SU(N)$ and $O(N)$ symmetries in the 
large $N$ limit, and using the spin stiffness $\rho$, as a parameter Read and
Sachdev obtained  results for $M(T)$ which are in reasonable good agreement
with the experimental data at low temperatures.
ii)Kasner and MacDonald\cite{kasner1} calculated 
$M(T)$ using many-particle
diagrammatic techniques which include spin-wave excitations
and electron spin-wave interaction.
This theory is a good improvement on the one-particle Hartree-Fock
theory, but it gives a polarization for the system too high compared
with the experimental one.
Progress in the 
diagrammatic approach, including 
temperature dependence  screening,
has been done by Haussman.\cite{haussman}
iii) The dependence of the magnetization on $T$ has been also obtained by exact 
diagonalization of the many-particle Hamiltonian for a small (up to 9)
number of electrons on a sphere.\cite{kasner2,chakraborty}
These calculation have important finite size effects at low temperature 
and weak  Zeeman coupling.
iv) Finally, 
quantum Monte Carlo (MC) techniques have been used in order to calculate  $M(T)$
for a spin 1/2 quantum Heisenberg model on a square lattice with exchange interactions 
adjusted to reproduce the spin stiffness of the
quantum Hall ferromagnet.\cite{henelius} These calculations are essentially 
exacts and
probably are free of finite size effects.

A unique property of the quantum Hall ferromagnets is the equivalence
between the topological charge associated with ${\bf m}$
and the electrical charge.\cite{libro1}
This equivalence make the Skyrmions to be the relevant charged excitation of
the 2DEG at $\nu=1$.
Charge conservation implies that at a given $\nu$
the integral of the topological charge over all the space
should be constant, independently of $T$. At $\nu=1$ this constant is zero.
A spin vector field texture produces a modulation of the topological
charge density.
Spatial spin fluctuations increase with temperature and produce
a modulation of the topological charge density.
In this way 
thermal
fluctuations can produce  a strong charge fluctuation.
Because of the equivalence between topological and real charge, the 
modulation of the charge density costs Hartree energy.
The models described above: diagrammatic techniques, quantum field theory
and quantum MC calculations, do not take into account the
Hartree contribution to the energy of the ferromagnet.

In this work we study the effect that the Hartree energy has on the temperature dependence
of the magnetization. We perform classical MC simulations of
$M(T)$ for the energy functional of the Hall ferromagnet. 
From the comparison of the results with and without Hartree energy
we conclude that at moderate temperatures the inclusion of this term
modifies the value of the magnetization up to a thirty per cent.
For realistic values of the Zeeman coupling,
we find that at intermediate-high temperature the Hartree energy is near
one third of the Zeeman and exchange energies.

The classical model can not describe correctly the low temperature behavior of $M(T)$.
The classical  dynamics of the electron spins
neglects several effects, in particular the quantum 
description of the spin-density waves
which is extremely important for describing $M(T)$ at low temperatures.
The inclusion,
in  the classical model 
of a temperature dependent low-energy cutoff
simulates quantum effects.
As we show latter,
in the Hall ferromagnet
this cutoff appears naturally 
when discretizing the continuum model.
In this work we are interested in  the effect that the 
inclusion of the Hartree term has on $M(T)$.
We expect that in the quantum Heisenberg model,
the Hartree energy should have the same effect than in the classical model.

The long-wavelength and low-energy properties of the $\nu=1$ Hall ferromagnet can be described
by a functional $E$ of the unit vector field  ${\bf m} ({\bf r})$
which  describes the local orientation of the spin magnetic order.
The functional $E$ has three terms\cite{libro1}, the gradient leading or
exchange  term $E_x$,
the Zeeman term $E_z$ and the Hartree term $E_c$.
\begin{eqnarray}
E & = & E_x+E_z+E_c \, \\
E_x & = & {{  \rho } \over 2} \int d ^2 r \left ( \nabla {\bf m} \right ) ^ 2
\\
E_z & = & 
 { {t} \over {2 \pi \ell ^2}} \int d ^2 r [ 1 - m _z({\bf r}) ] \, \, \,  \\
E_c & = &     {1 \over 2} \int d ^2 r d ^2 r '
n({\bf r}) n({\bf r}') v (|{\bf r} - {\bf r '}|) \, \, \, \, \, ,
\end{eqnarray}
here $t= g ^* \mu _B B /2$ is the Zeeman coupling strength, 
$\ell$ is the 
magnetic length, $v (|{\bf r} - {\bf r '}|)$ is the Coulomb interaction
and $n({\bf r})$ is the charge density and it is given by\cite{rajaraman}
\begin{equation}
n({\bf r})= { 1 \over {8 \pi}} \epsilon _ {\nu \mu}  \, {\bf m} ({\bf r}) \cdot
\left [ \partial _ {\nu} {\bf m} ({\bf r}) \times
\partial _ {\mu} {\bf m} ({\bf r}) \right ]
\, \, \, \, \,  .
\end{equation}
In this  continuum model the total topological charge, $Q$, is given by the integral over all
the space of $n ({\bf r})$, and it represents the number of times 
${\bf m} ({\bf r})$ winds around the sphere $S^2$.
Skyrmions are non trivial extrema solutions of  the 
functional $E$ with $Q\neq0$.
The continuum model gives a good description of the Skyrmions with moderate 
and large spins.\cite{abolfath}

In order to perform classical MC calculations we discretize the space using
a square lattice with a lattice parameter $a_L$.
The lattice parameter corresponding
to an electron per unit cell is 
$a_1 = \sqrt{2 \pi} \ell$.
The  energy functional has the form,
\begin{equation}
E \!  = \!  - \rho \!\! \sum _ {<i,j>} {\bf \Omega} _i  \, {\bf \Omega} _j \!
 -t 
{{ a _L ^2} \over {a_1 ^ 2}}
\! \sum _ i \left ( \Omega _{z,i} +1 \! \right ) 
\! + \! {1 \over 2} \! \sum _{i,j} q _i \, V _{i,j} \, q_j   .
\end{equation}
Here, ${\bf \Omega} _i$ is the unit vector at site $i$, 
$q _i $ is the topological charge attached
to the unit cell $i$ and 
$V _{i,j} $ is the
Coulomb interaction between two charges unity  distributed uniformly in the 
cells $i$ and $j$. 
The cell $i$ is defined by the points 1:($i_x$,$i_y$), 
2:($i_x$ +1 ,$i_y$), 
3:($i_x$ +1,$i_y$+1) and  
4:($i_x$,$i_y$ +1).
The expression of $q_i$ as a function of the unit vectors at the points 1-4 is\cite{berg} 
\begin{equation}
q_i = {1 \over {4 \pi}} \left \{
(\sigma A )
({\bf \Omega} _1,
{\bf \Omega} _2,
{\bf \Omega} _3) + 
(\sigma A )
({\bf \Omega} _1,
{\bf \Omega} _3,
{\bf \Omega} _4)   \right    \} \, \, \, \, ,
\end{equation}
where $(\sigma A )
({\bf \Omega} _1,
{\bf \Omega} _2,
{\bf \Omega} _3) $ denoted the signed area of the spherical
triangle with corners  
${\bf \Omega} _1,
{\bf \Omega} _2$ and 
${\bf \Omega} _3 $. 
Apart from a set of 'exceptional' configurations of measure zero, 
this prescription for the topological charge
yields well defined integer values for the total topological charge.
For smooth spin texture the continuum and discrete expression for the density of
topological charge give the same value. 
In order to analyze the effect of the Hartree term we study also the
functional $E_0=E_x+E_z$,
that 
is the classical version
of the quantum Heisenberg Hamiltonian studied by Henelius {\it et al.}\cite{henelius}
In that work the Hartree term is not taken into account.

A comment about the significance of $a_L$ is in order.
In two dimensions the exchange term and the topological charge
are  scale invariants 
and do not depend on $a_L$. The Zeeman energy increases quadratically
with the lattice parameter and  $V_{i,j}$ is inversely proportional
to $a_L$. 
An increment in $a_L$ is similar to  an increment  of the Zeeman strength
and the lattice parameter acts as a low energy cutoff that 
controls the dynamics of the classical Heisenberg model.
We obtain the value of $a_L$ by fitting the magnetization
obtained from the functional $E_0$ to the magnetization obtained  from quantum MC simulations.\cite{henelius}
In this way we obtain a temperature dependent $a_L$. 
In the $T \rightarrow \infty$ limit $a_L$ should be $a_1$.

Now we describe briefly the MC procedure used for obtaining $M(T)$.
The MC simulations were performed
by using the techniques due to 
Metropolis {\it et al.}\cite{metropolis}
we study a cluster $N \times N$ with periodic boundary conditions (PBC).
The use of PBC diminish the finite size effects.
For studying $M(T)$ at $\nu=1$ we consider as a starting configuration
a completely ordered ferromagnet, i.e. ${\bf \Omega } _i = (0,0,1)$ for
all sites $i$. In this way the 
initial $Q$ in the system is zero.
The sites to be considered for a change in the spin orientation
are randomly chosen, avoiding artificial correlations that
could distort the results.
Once a site is selected for a spin    reorientation,
we perform the following operations:
i) a small change in the direction of ${\bf \Omega} _i$,
ii)calculation of the change induced in  the topological
charge by the spin reorientation.
Only changes that do not modify $Q$ are accepted.
iii)evaluation of the energy variation $\Delta E$.
iv) acceptance or not acceptance of the new spin direction. If
$\Delta E <0$ the change is accepted. If $\Delta E >0$
the change is accepted depending if $e^{-\Delta E/kT}$ exceeds a random number.
We realize a large number of MC steps until reach the equilibrium situation
and perform the average of the different statistical properties:
the magnetization $M(T)$,
the different contributions  to the total energy per electron, $<E_x>$, $<E_z>$ and $<E_c >$
and a measure of the charge fluctuations,
\begin{equation}
\delta q=\sqrt{ { {\sum _i q _i ^ 2 } \over {N \times N}}} \, \, \, .
\end{equation}

In Fig.1 we plot the magnetization as a function of temperature as obtained from the
classical MC simulation for the functional $E_0=E_x+E_z $
with different value of the lattice parameter. 
The results correspond to a 2DEG of zero layer thickness, $\rho = 0.0249 e^2 / \ell$
and a Zeeman coupling $t=0.008 e ^2 / \ell$.
The calculations are performed in a cluster 20$\times$20. We have checked that
for this cluster size and PBC, the results are free of size effects.
In the same figure we plot $M(T)$ as obtained from quantum MC
simulations.\cite{henelius}.
This calculation does not include the Hartree term and it is the quantum version
of the functional $E_0$.
By comparing the classical and the quantum results we can estimate 
$a_L$ in the different  temperature ranges. We are interested in temperatures
in the range  $0.075e^2\ell< T<0.2 e ^2 / \ell$. At lower temperatures
the quantum effects are very important and at higher temperatures
the effective functional Eq.(1) it is not longer valid.
In order to describe the dynamic in this range of temperatures  
values of $a_L$ in the range $2 a_1 < a_L < 2.5 a_1$ are  necessary.
For simplicity, in the calculation we use a constant value of the lattice parameter.
We use 
$a_L$=2.5$a_1$, which is the value of $a_L$ in the range 
$2 a_1 < a_L < 2.5 a_1$ where the effects of the Hartree interaction are weaker.

In Fig.2 we plot $M(T)$ as obtained from the classical MC simulation for the
full functional $E=E_x+E_z+E_c$ and for the functional without Hartree
energy $E_0=E_x+E_z$. the results corresponds to a lattice parameter 
$a_L=2.5 a_1$.
At low temperatures, $T<0.05 e ^2 / \ell$, the spins are not very disordered, the
charge modulation is very weak and therefore the Hartree 
term has a small effect on $M(T)$. For higher temperatures the spin fluctuations
and consequently the charge fluctuations are stronger and  
the  Hartree term becomes a important contribution to the internal energy.
We find that at intermediate temperatures, 
in the  range $0.05 e ^ 2/\ell < T < 0.15 e ^2 \ell$ the magnetization
obtained with the full functional  $E$ is near thirty per cent 
higher than 
the obtained without
the Hartree term. At  even higher temperatures,
$T > 0.15 e ^2 / \ell$, the spin disorder is very large and the magnetization
calculated with 
or without Hartree term is small, although $M(T)$ obtained with $E$ is 
always higher than the obtained with $E_0$.

As commented above the quantum MC calculation describe correctly  
the spin density waves and at low temperatures it should give an appropriate
$M(T)$
for the $\nu =1$ Hall ferromagnet. However we expect that for intermediate temperatures
the Hartree energy term would modify the quantum MC data
in a similar amount that it modifies  the classical results.
In order to understand the experimental results at intermediate temperatures 
it is necessary to take into account
the charge fluctuations induced by the temperature.

In figure 3 we plot the different contributions to the total 
energy per electron as a function of the temperature.
The parameters are the same than the used in figure 2. At very high
temperatures the spin are completely random and the exchange and Zeeman 
energies tend to their fully disorder values $2\rho / a_L ^ 2 $ and $t$ respectively.
Note that intermediate temperatures the Hartree energy is near one third
the Zeeman energy. For smaller values of the Zeeman coupling  and high temperatures 
we have found that the Hartree energy can be the more important 
contribution to the internal energy.

Figure 4 shows the variation of the charge fluctuations, $\delta q$ 
as a function of the temperature.
Charge fluctuations cost Hartree energy and they  are weaker when 
the full functional is considered. Observe that 
at intermediate temperatures $\delta q$ is of the order of 0.1, that
is $\sim 10\%$ of the charge in each cell.

In closing, we have studied the effect that the Hartree energy term has
on the temperature dependence of a 2DEG at $\nu=1$. We find
that at intermediate temperatures the spin fluctuations are  weakened
by the Hartree energy and the magnetization is near thirty per cent
bigger 
than the obtained by neglecting the Hartree energy term. At intermediate
temperatures the Hartree energy is an important contribution to the 
internalñ energy of the Hall ferromagnet.

We thank A.H.MacDonald, C.Tejedor, L.Mart\'{\i}n-Moreno and J.J.Palacios 
for useful discussions.
This work was supported by the CICyT of Spain under Contract No. PB96-0085
and by  the
Fundaci\'on Ram\'on Areces.

\vspace{-6.8truecm}
\begin{figure}
\epsfig{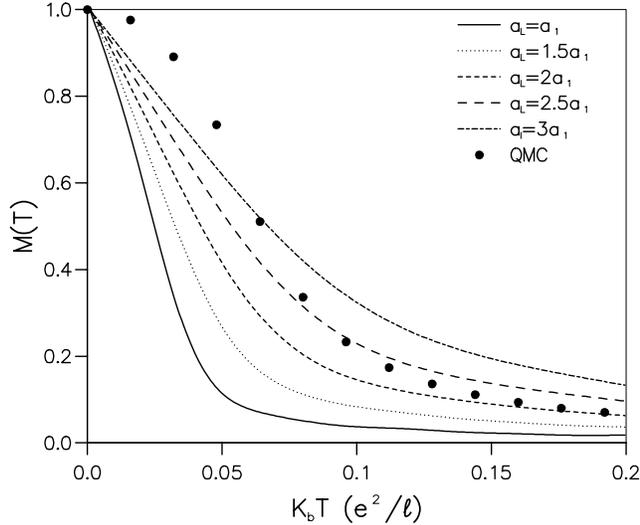}
\caption{
M(T) as obtained by using the functional $E_0$   with
different lattice parameters. We also plot the quantum MC
(QMC) results.[13]
The results correspond to a 2DEG of zero layer thickness
and a Zeeman coupling $t=0.008 e ^2/\ell$.
}
\end{figure}

\vspace{-7.0cm}
\begin{figure}
\epsfig{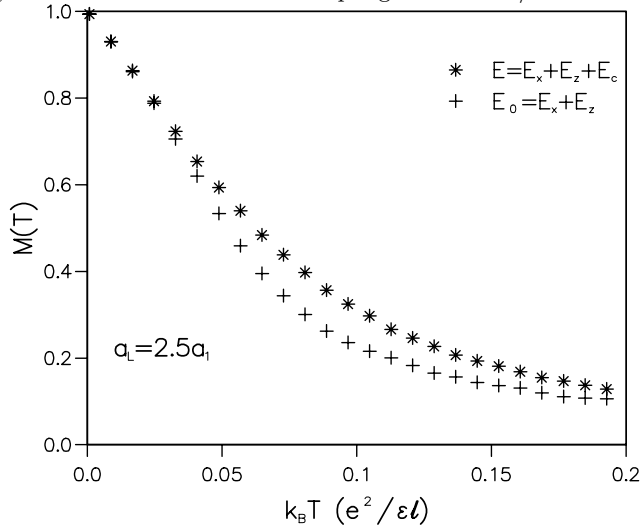}
\caption{
M(T) as obtained by using the functional $E$   and
as obtained by neglecting the Hartree term. 
The results correspond to a lattice parameter $a_L =2.5 \sqrt{2 \pi}\ell$.
}
\end{figure}
\newpage
.
\vspace{-6truecm}
\begin{figure}
\epsfig{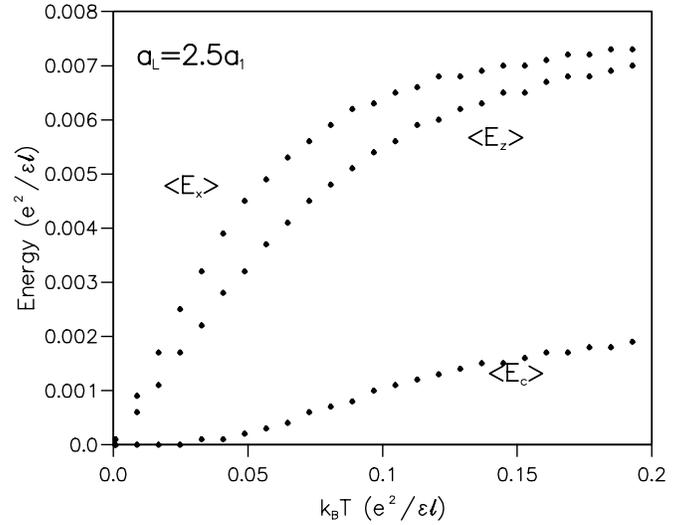}
\caption{
Different contributions to the internal energies per electron. 
}
\end{figure}
\vspace{-5cm}
\begin{figure}
\epsfig{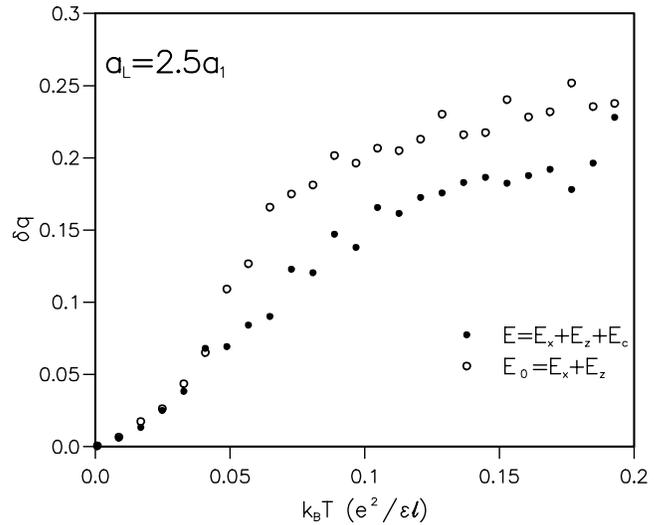}
\caption{
Variation of $\delta q$, equation (8), as a function of the temperature
for the full functional $E$ and for the functional without Hartree energy $E_0$.
}
\end{figure}

\end{document}